\documentclass[aip,apl,reprint]{revtex4-1}
\usepackage{amsmath}
\usepackage{dcolumn}
\usepackage{bm}
\usepackage{graphicx}
\usepackage{amsfonts}
\usepackage{amssymb}
\usepackage{bbm}
\usepackage{float}

\begin{document}

\title{Quantum-memory-assisted entropic uncertainty relation with a single
nitrogen-vacancy center in diamond}

\begin{abstract}
The limitation of simultaneous measurements of noncommuting observables can
be eliminated when the measured particle is maximally entangled with a
quantum memory. We present a proposal for testing this
quantum-memory-assisted entropic uncertainty relation in a single
nitrogen-vacancy (N-V) center in diamond only by local electronic
measurements. As an application, this entropic uncertainty relation is used
to witness entanglement between the electron and nuclear spins of the N-V
center, which is close to reach the currently available technology.
\end{abstract}

\author{Z. Y. Xu $^{1}$}
\email{zhenyuxu@suda.edu.cn}
\author{S. Q. Zhu $^{1}$}
\email{szhu@suda.edu.cn}
\author{W. L. Yang $^{2}$}

\affiliation{$^{1}$School of Physical Science and Technology, Soochow University, Suzhou,
215006, China \\
$^{2}$State Key Laboratory of Magnetic Resonance and Atomic and Molecular
Physics, Wuhan Institute of Physics and Mathematics, Chinese Academy of
Sciences, Wuhan, 430071, China}
\maketitle

The Heisenberg's uncertainty principle is the cornerstone of quantum
mechanics, which provides a limitation of simultaneous measurements of
canonically conjugate observables \cite{UP}. In quantum information theory,
the uncertainty relation is usually characterized by entropic measures
rather than standard deviations \cite{UP-Review0,UP-Review}. However, the
traditional entropic uncertainty relation may be violated if particle $A$ to
be measured is initially entangled with a quantum memory $B$ \cite{violation}%
. Recently, a quantum-memory-assisted entropic uncertainty relation has been
conjectured \cite{eEUP-0} and later proved with an equivalent form \cite%
{eEUP}
\begin{equation}
S\left( Q|B\right) +S\left( R|B\right) \geq \log _{2}\frac{1}{c}+S\left(
A|B\right) .
\end{equation}%
Here, $S\left( \Lambda |B\right) $ with $\Lambda \in (Q,R)$ is the
conditional von Neumann entropy of the post-measurement state $\rho
_{\Lambda B}=\sum_{m}(|\psi _{m}\rangle \left\langle \psi _{m}\right\vert
\otimes \mathbbm{1})\rho _{AB}(|\psi _{m}\rangle \left\langle \psi
_{m}\right\vert \otimes \mathbbm{1}),$ where $\mathbbm{1}$ is the identity
operator and \{$|\psi _{m}\rangle $\} are the eigenstates of the observable $%
\Lambda $. The parameter $\log _{2}\frac{1}{c}$ quantifies the
incompatibility of two measurements with $c=$ max$_{\alpha ,\beta
}\left\vert \left\langle \phi _{\alpha }\right\vert \varphi _{\beta }\rangle
\right\vert ^{2}$, where $|\phi _{\alpha }\rangle $ and $|\varphi _{\beta
}\rangle $ are the eigenstates of two noncommuting observables $Q$ and $R$,
respectively. Equation (1) can be understood in a straightforward way: the
entanglement between $A$ and $B$ may create a negative conditional entropy $%
S(A|B)$ \cite{Book-QIC}, which will beat the incompatibility $\log _{2}\frac{%
1}{c}$. In other words, the quantum information held by the quantum memory $%
B $ will help us to eliminate the uncertainty of the measurements performed
on particle $A$. This entropic uncertainty relation together with other
forms \cite{Other-measures,Other-measures2,lieb} has practical applications
such as for witnessing entanglement \cite{eEUP} and cryptography \cite{MIMA}
and has been explored under noise \cite{xzy} and in optical photon systems
\cite{exp1,exp2}.

Over the past few years, the negatively charged nitrogen-vacancy (N-V)
center in diamond has been considered as one of the most promising building
block for solid-state quantum information processing \cite{NV-Review}.
Indeed, due to weak magnetic interaction with the environment, the proximal
nuclear spins in a single N-V center become a good quantum memory for
information storage \cite{NV-memory2007,NV-memory2011,NV-memory2012}. In
addition, the paramagnetic electron spin in a single N-V center can be
optically polarized and readout with high fidelity at room temperature \cite%
{NV-review2} and can also be used to polarize and readout the nuclear spins
\cite{NV-single-shot-nuclear} and realize conditional gating \cite%
{NV-CNOT,NV-CNOT2012}. Based on above advantages, the N-V center provides an
excellent test bed to explore this quantum-memory-assisted entropic
uncertainty relation.

In this Letter, we present a practical proposal to verify this entropic
uncertainty relation in a single N-V center in diamond by local electronic
measurements. As a by-product, this entropic uncertainty relation is
employed to witness entanglement of electron and nuclear spins in diamond.
The experimental feasibility is also justified with current laboratory
parameters.

\begin{figure}[tbp]
\centering\includegraphics[width=7 cm]{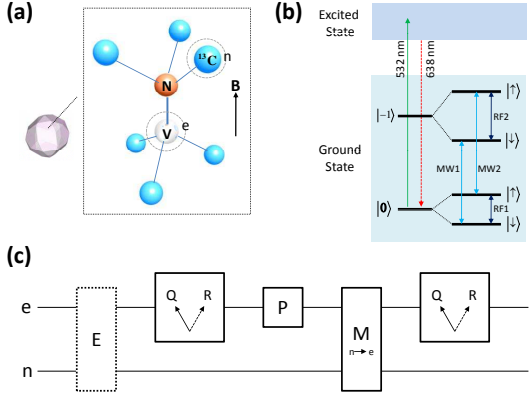}
\caption{(Color online) (a) Structure of a single N-V center in diamond
including the electron spin and the $^{13}$C nuclear spin. An external
magnetic field $\mathbf{B}$ is applied along the [111] axis of the N-V
center to form the quantization axis. (b) Electron--nuclear spin level
diagram, where electronic ground states $\{\left\vert 0\right\rangle
,\left\vert -1\right\rangle \}$ and nuclear spin states $\{\left\vert
\downarrow \right\rangle ,\left\vert \uparrow \right\rangle \}$ are employed
to encode the two-qubit system. (c) Schematics for testing the
quantum-memory-assisted entropic uncertainty relation and witnessing the
entanglement between electron and nuclear spins. The first dashed box is
only for testing the uncertainty relation, which is not necessary for the
procedures of witnessing entanglemant. $Q=\protect\sigma _{1}$ and $R=%
\protect\sigma _{3}$ in the second box are two noncommuting observables. The
third box represents the polarization of electron spin and the forth box is
used to transfer the nuclear state to the electronic state by mapping
operations, which will be followed by the fifth box by performing the same
measurement as the second box on the electron spin again.}
\end{figure}

Before introducing the experimental proposal, we first display a theoretical
framework of the quantum-memory-assisted entropic uncertainty relation for
two-qubit system, in which an arbitrary two-qubit state can be of the form
\cite{entanglement} $\rho _{AB}=\frac{1}{4}[\mathbbm{1}^{A}\otimes %
\mathbbm{1}^{B}+\sum_{i=1}^{3}\left( x_{i}\sigma _{i}^{A}\otimes \mathbbm{1}%
^{B}\mathbf{+}\mathbbm{1}^{A}\otimes y_{i}\sigma _{i}^{B}\right)
+\sum_{i,j=1}^{3}T_{ij}\sigma _{i}^{A}\otimes \sigma _{j}^{B}]$, where $%
\sigma _{i(j)}$ with $i(j)\in \{1,2,3\}$ correspond to standard Pauli
matrices. We define two $3\times 1$ vectors $\mathbf{x}$ with real
components $x_{i}=$tr$_{AB}(\rho _{AB}\sigma _{i}^{A}\otimes \mathbbm{1}^{B})
$ and $\mathbf{y}$ with real components $y_{i}=$tr$_{AB}(\rho _{AB}%
\mathbbm{1}^{A}\otimes \sigma _{i}^{B}),$ and a $3\times 3$ correlation
tensor $\mathbf{\bar{T}}$ with real components $T_{ij}=$tr$_{AB}(\rho
_{AB}\sigma _{i}^{A}\otimes \sigma _{j}^{B})$. If we choose two Pauli
observables $Q=\sigma _{1}$ and $R=\sigma _{3}$ to be measured, the
uncertainty, i.e., the left hand side of Eq. (1), can be expressed as
\begin{equation}
U=-\sum_{\substack{ \mu ,\nu =0,1 \\ \lambda =1,3}}\eta _{\mu \nu }^{\lambda
}\log _{2}(\eta _{\mu \nu }^{\lambda })-2H_{bin}\left( \frac{1-\left\Vert
\mathbf{y}\right\Vert }{2}\right) ,
\end{equation}%
where $H_{bin}\left( p\right) =-p\log _{2}p-(1-p)\log _{2}(1-p)$ denotes the
binary entropy \cite{Book-QIC}, $\eta _{\mu \nu }^{\lambda }=[1+(-1)^{\mu
}x_{\lambda }+(-1)^{\nu }\sqrt{\sum_{i=1}^{3}(y_{i}+(-1)^{\mu }T_{\lambda
i})^{2}}]/4,$ and $\left\Vert \mathbf{y}\right\Vert =\sqrt{%
\sum_{i=1}^{3}y_{i}^{2}}$. Since the complementarity $c$ of the observables $%
\sigma _{1}$ and $\sigma _{3}$ is always equal to 1/2, the lower bound of
uncertainty, i.e., the right hand side of Eq. (1), takes the form
\begin{equation}
U_{b}=S\left( \rho _{AB}\right) +1-H_{bin}\left( \frac{1-\left\Vert \mathbf{y%
}\right\Vert }{2}\right) .
\end{equation}

In experiment, if we choose the same measurement $\Lambda $ on particles $A$
and $B$, we may get $H(\Lambda |\Lambda )\geq S\left( \Lambda |B\right) $
with $H\left( \cdot \right) $ the Shannon entropy. According to Fano's
inequality \cite{Book-QIC}, we have $H(\Lambda |\Lambda )\leq H_{bin}\left(
\kappa _{\Lambda }\right) $ with $\kappa _{\Lambda }$ the probability that
the outcomes of the same measurement $\Lambda $ on particles $A$ and $B$ are
different. Therefore, $H_{bin}\left( \kappa _{\Lambda }\right) \geq S\left(
\Lambda |B\right) ,$ which will in general lead to a higher measurement
estimation ($ME$) of the uncertainty, and can be used to conveniently test $U
$ with experimental counts \cite{exp1,exp2}. For two-qubit states and
observables $\sigma _{1}$ and $\sigma _{3}$,
\begin{equation}
ME=H_{bin}\left( \frac{1-T_{11}}{2}\right) +H_{bin}\left( \frac{1-T_{33}}{2}%
\right) .
\end{equation}

In Fig. 1, the structure of a pure N-V center in diamond is depicted, where
we treat the electron spin ($S=1)$ as particle $A$ and a proximal $^{13}$C
nuclear spin ($I=1/2)$ as quantum memory $B$. In the following, we will
employ the electronic states $\{\left\vert 0\right\rangle ,\left\vert
-1\right\rangle \}$ and the nuclear states $\{\left\vert \downarrow
\right\rangle ,\left\vert \uparrow \right\rangle \}$ as two-qubit system
[Fig. 1(b)].

An experimental schematic sequence for testing this quantum-memory-assisted
entropic uncertainty relation is depicted in Fig. 1(c). The first dashed box
represents the entangled state (for example, we consider the Schmidt state $%
\left\vert \Phi \right\rangle =\cos \chi \left\vert 0\downarrow
\right\rangle +\sin \chi \left\vert -1\uparrow \right\rangle $ with $\chi
\in \lbrack 0,\pi /2]$) preparation with the following key steps \cite%
{NV-memory2007}: (i) The electronic ground state is first polarized at state
$\left\vert 0\right\rangle $ by a \textit{532} nm green laser pulse and the
two-qubit state now reads $\left\vert 0\right\rangle \left\langle
0\right\vert \otimes \rho _{n}$ with $\rho _{n}$ an unknown nuclear mixed
state; (ii) Then we can transfer the nuclear state to the electronic state $%
\left\vert 0\right\rangle \left\langle 0\right\vert \otimes \rho
_{n}\rightarrow $ $\rho _{e}\otimes \left\vert \downarrow \right\rangle
\left\langle \downarrow \right\vert $ with a conditional MW1-$\pi $ pulse
followed by a conditional RF1-$\pi $ pulse; (iii) The electronic state can
then be polarized again by the \textit{532} nm laser pulse, which reduces
the two-qubit state to $\left\vert 0\right\rangle \left\vert \downarrow
\right\rangle $; (iv) A MW1-2$\chi $ pulse is then performed on the
electronic state, which yields the product state $(\cos \chi \left\vert
0\right\rangle +\sin \chi \left\vert -1\right\rangle )\otimes \left\vert
\downarrow \right\rangle $. Then a conditional RF2-$\pi $ pulse will create
the final entangled state $\left\vert \Phi \right\rangle .$

After preparation of the entangled state, we then perform measurement $%
Q=\sigma _{1}($or $R=\sigma _{3})$ on the electron spin. In our case, the $%
\sigma _{3}$ operation can be directly manipulated by electron shelving
projecting onto $\left\vert 0\right\rangle $ or $\left\vert -1\right\rangle $
with \textit{532} nm laser pulse, and the $\sigma _{1}$ operation can be
reduced to detecting $\sigma _{3}$ by applying a stronger MW-$\pi /2$ pulse
before and after $\sigma _{3}$ operation (i.e., $\sigma _{1}=\mathcal{H}%
\sigma _{3}\mathcal{H}$ with $\mathcal{H}$ the Hadamard gate).

\begin{figure}[tbp]
\centering\includegraphics[width=7 cm]{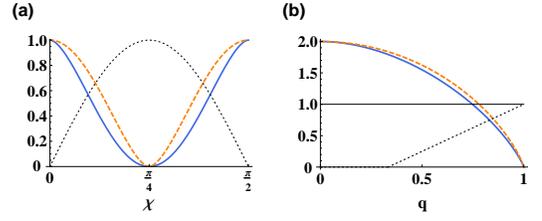}
\caption{(Color online) (a) An example of testing the
quantum-memory-assisted entropic uncertainty relation with Schmidt state $%
\left\vert \Phi \right\rangle =\cos \protect\chi \left\vert 0\downarrow
\right\rangle +\sin \protect\chi \left\vert -1\uparrow \right\rangle $ and
observables $Q=\protect\sigma _{1}$ and $R=\protect\sigma _{3}$. The
uncertainty $(U=U_{b})$, measurement estimation $(ME)$, and the concurrence $%
(C)$ are represented by blue solid, orange dashed, and black dotted curves
respectively. (b) Entanglement witness of electron and nuclear spins in a
single N-V center. The blue solid, orange dashed, and black dotted curves
denotes $U_{b}$, $U=ME$, and $C$, respectively. }
\end{figure}

To acquire the measurement estimation of the uncertainty in experiment, the
same observable $Q=\sigma _{1}($or $R=\sigma _{3})$ should be performed on
the proximal $^{13}C$ nuclear spin. However, since electron spin is good for
efficient processing and readout, while the nuclear spin is more suitable
for long-term storage \cite{NV-en}, we will transfer the nuclear state to
the electron spin and then perform the same operation $\sigma _{1}$ $($or $%
\sigma _{3})$ on the electron spin to quantify the uncertainty. This task
can be accomplished by first polarizing the electron spin to the state $%
\left\vert 0\right\rangle $ again [the 3rd box in Fig. 1(c)], and then
employing the conditional MW2-$\pi $ pulse followed by a conditional RF2-$%
\pi $ pulse to transfer the nuclear state to the electronic state [the 4th
box in Fig. 1(c)]. Finally, we may perform the $\sigma _{1}$ $($or $\sigma
_{3})$ operation on the electron spin again [the 5th box in Fig. 1(c)]. In
comparison with the first and the second outcomes of $\sigma _{1}$ $($or $%
\sigma _{3})$ operation [the 2nd and 5th boxes in Fig. 1(c)], we may acquire
the measurement estimation of the uncertainty.

For state $\left\vert \Phi \right\rangle $ with $\mathbf{x=y=}(0,0,\cos
2\chi )^{t}$ and $\mathbf{\bar{T}=}diag\{\sin 2\chi ,-\sin 2\chi ,1\}$, it
is easy to check that the uncertainty $U=U_{b}=1-H_{bin}\left( \sin ^{2}\chi
\right) $ and the measurement estimation $ME=H_{bin}\left( \frac{1-\sin
2\chi }{2}\right) $. In Fig. 2(a), the dependence of the uncertainty $U$ and
the measurement estimation $ME$ on the Schmidt state angle $\chi $ is
plotted. The blue solid line represents the uncertainty $U$ and the lower
bound $U_{b}$ (the equality is achieved in this case). The orange dashed
line represents the measurement estimation, which is higher than the
uncertainty$.$ To illustrate the role of entanglement on this
quantum-memory-assisted entropic uncertainty relation, we employ the
concurrence \cite{concurrence} with $C=2max\{0,\left\vert \sin \chi \cos
\chi \right\vert \}$ [black dotted line in Fig. 2(a)] to characterize the
entanglement. It is clearly illustrated that when the electron and nuclear
spins are in a maximum entangled state ($\chi =\pi /4$), the uncertainty of
two noncommuting observables can be totally eliminated.

\begin{figure}[tbp]
\centering\includegraphics[width=4 cm]{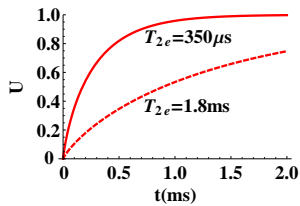}
\caption{(Color online) Electronic dephasing effect on the uncertainty $%
U(=U_{b}=ME)$ of observables $Q=\protect\sigma _{1}$ and $R=\protect\sigma %
_{3}$ performed on the state $\left\vert \Phi ^{+}\right\rangle =\frac{1}{%
\protect\sqrt{2}}(\left\vert 0\downarrow \right\rangle +\left\vert
-1\uparrow \right\rangle ).$ Parameters are adopted in Ref. \protect\cite%
{NV-e-T2-350} with $T_{2e}=350$ $\protect\mu s$ and in Ref. \protect\cite%
{NV-e-T2-1.8ms} with $T_{2e}=1.8$ $ms.$}
\end{figure}

One of the most remarkable applications of this quantum-memory-assisted
entropic uncertainty relation is to witness entanglement of electron and
nuclear spins in the N-V center. Our scheme with only local electronic
measurements and no need for quantum state tomography is helpful for
experimental implementation. The procedures are similar to the former
testing case except for the initial state preparation, i.e., the preparation
of initial entangled state is actually not required, so the first box in
Fig. 1(c) could be removed. However, to illustrate the application on the
entanglement witness, we assume that the electron and nuclear spins are in
state $\left\vert \Xi \right\rangle =\frac{1-q}{4}\mathbbm{1}\otimes %
\mathbbm{1}+q\left\vert \Phi ^{+}\right\rangle \left\langle \Phi
^{+}\right\vert ,$ with $\left\vert \Phi ^{+}\right\rangle =\frac{1}{\sqrt{2}%
}\left\vert 0\downarrow \right\rangle +\left\vert -1\uparrow \right\rangle .$
For state $\left\vert \Xi \right\rangle $ with $\mathbf{x=y=}(0,0,0)^{t}$
and $\mathbf{\bar{T}=}diag\{q,-q,q\}$, we have $U=ME=2H_{bin}\left( \frac{1-q%
}{2}\right) $, $U_{b}=H_{bin}\left( \frac{1+3q}{4}\right) +\frac{3(1-q)}{4}%
\log _{2}3$, and $C=2\max \{0,\frac{3q-1}{4}\}.$ As shown in Fig. 2(b), the
electron and nuclear spins must be entangled if the uncertainty is less than
1.

Finally, we survey the relevant experimental parameters. As reported in
recent N-V experiments, the longitudinal relaxation time $T_{1n}$ of $^{13}C$
nuclear spin is around $1.7$ $s$ \cite{NV-memory2012}, and the coherence
time $T_{2n}$ is about $20$ $ms$ \cite{NV-memory2007} (We note that $T_{2n}$
of $^{13}C$ can reach on the order of one second with elegant dissipative
decoupling technique (DDT) \cite{NV-memory2012}. However, the straight
forward employment of DDT in our scheme will also induce complications with
ionization and deionization of the N-V center). While for N-V electron spin
at room temperature, the relaxation time $T_{1e}=6$ $ms$ is reported in Ref.
\cite{NV-e-T1-6ms} and coherence time $T_{2e}$ ranges from from $350$ $\mu s$
\cite{NV-e-T2-350} to $1.8$ $ms$ \cite{NV-e-T2-1.8ms}. Since the key step of
our proposal is the state mapping from nucleus to electron, which will take
on the time scale of 100 $\mu s$ \cite{NV-CNOT2012}, $T_{2e}$ becomes the
key factor in our proposal ($T_{2n}=20$ $ms$ \cite{NV-memory2007} is long
enough for implementing our proposal). Here, we briefly study the electronic
dephasing effect on this entropic uncertainty relation. As an example, the
initial state $\left\vert \Phi ^{+}\right\rangle =\frac{1}{\sqrt{2}}%
(\left\vert 0\downarrow \right\rangle +\left\vert -1\uparrow \right\rangle )$
under electronic dephasing effect yields $U=U_{b}=ME=H_{bin}\left( \frac{%
1-e^{-\frac{t}{2T_{2e}}}}{2}\right) $, which is depicted in Fig. 3. Clearly,
longer electronic dephasing time $T_{2e}$ will retain less uncertainty of
two incompatible measurements.

In conclusion, the quantum-memory-assisted entropic uncertainty relation has
been explored in a single N-V center in diamond. By investigating relevant
experimental parameters, our proposal can be immediately verified under
current N-V-based experimental conditions\emph{.}

This work is supported by NNSFC under Grant Nos. 11204196, 11074184,
11004226, and 11274351.

\end{document}